\documentclass[]{spie}  

\usepackage{amsmath,amsfonts,amssymb}
\usepackage{graphicx}
\usepackage[colorlinks=true, allcolors=blue]{hyperref}
\usepackage[flushleft]{threeparttable}

\title{A PyTorch Benchmark for High-Contrast Imaging Post Processing}

\author[a]{Chia-Lin Ko}
\author[a]{Ewan S. Douglas}
\author[a]{Justin Hom}
\affil[a]{Department of Astronomy and Steward Observatory, University of Arizona, Tucson, AZ, USA}

\begin{document} 
\maketitle

\begin{abstract}
Direct imaging of exoplanets is a challenging task that involves distinguishing faint planetary signals from the overpowering glare of their host stars, often obscured by time-varying stellar noise known as ``speckles". The predominant algorithms for speckle noise subtraction employ principal-based point spread function (PSF) fitting techniques to discern planetary signals from stellar speckle noise. We introduce \texttt{torchKLIP}, a benchmark package developed within the machine learning (ML) framework \texttt{PyTorch}.  This work enables ML techniques to utilize extensive PSF libraries to enhance direct imaging post-processing. Such advancements promise to improve the post-processing of high-contrast images from leading-edge astronomical instruments like the James Webb Space Telescope and extreme adaptive optics systems. 
\end{abstract}


\section{INTRODUCTION}
\label{sec:intro}  

Exoplanets, planets located beyond our Solar System, hold the key to understanding our place in the Universe. Detecting and characterizing exoplanets is critical for advancing our knowledge of planetary systems. There are various methods to detect exoplanets\cite{Fischer2014}, such as with radial velocity, transit, microlensing, and direct imaging measurements. Among these, direct imaging stands out as it provides direct evidence of exoplanets, offering insights into mass, orbital periods, and atmospheric composition that are often elusive with indirect methods\cite{Currie2023}. 
However, direct imaging is also one of the most challenging techniques in exoplanet detection. While over 7000 exoplanets\footnote{\url{https://exoplanet.eu}} have been confirmed to date, fewer than 70 have been directly imaged. 
This challenge arises because the host stars are significantly brighter than the faint light emitted by their orbiting planets.
The contrast in brightness between the planet and starlight can be staggering, with differences reaching up to $10^{-6}$ for young Jovian planets, and as extreme as $10^{-9}$ for Jupiter-sized planets and $10^{-10}$ for Earth-sized ones\cite{Follette2023}. 

To enhance our ability to detect these faint exoplanets in high-contrast imaging (HCI), we utilize the support of coronagraph instruments, adaptive optics (AO) techniques\cite{Milli2016}, and advanced data processing algorithms. 
Coronagraphs are essential for blocking out the overwhelming light from the host star, while adaptive optics are critical for correcting atmospheric distortions. 
Despite advancements in coronagraph and adaptive optics technology, speckle noise—manifested as fluctuations in the host star's point spread function (PSF)—remains a significant challenge, often being mistaken for a planetary signal. These fluctuations in the stellar PSF are induced by varying atmospheric conditions and instabilities in instrument performance over time. To mitigate this issue effectively, specific observational strategies coupled with post-processing techniques are essential. A well-established post-processing technique is the Karhunen-Loève Image Projection (KLIP)\cite{Soummer2012}, a principal component analysis (PCA)\cite{Jolliffe2016}-based method for PSF subtraction. These approaches are crucial for accurately distinguishing between instrument-induced speckle noise and authentic planetary signals, as will be explored in more detail in Section \ref{methodology}.

However, even with the application of current data processing algorithms, the achievable contrast ratio is limited to around $10^{-5}$ to $10^{-6}$\cite{Guyon2005}, which is insufficient for detecting Earth-like exoplanets that require a contrast ratio of $10^{-10}$. Achieving such deep contrast ratios may depend on future telescopes, such as the proposed Habitable Worlds Observatory (HWO)\cite{Gaudi2020}, expected to launch in the 2030s, which will be equipped with superior instruments capable of reaching these levels. In the meantime, continued efforts to develop more effective data post-processing algorithms are crucial. Recent studies suggest that integrating machine learning methods into post-processing algorithms could significantly enhance our capabilities for exoplanet detection\cite{Cantalloube2020, Gebhard2022, Flasseur2024, Bonse2024}.

To enhance data post-processing capabilities, this paper aims to develop a tool called \texttt{torchKLIP}, designed to implement the basic KLIP algorithm within a machine learning (ML) framework using \texttt{PyTorch}\cite{Paszke2019}. \texttt{PyTorch} was chosen for its flexibility and potential for future integration with ML and deep learning (DL) algorithms. 
Existing data post-processing packages, such as \texttt{pyKLIP} \cite{Wang2015}, VIP\cite{Gonzalez2017, Christiaens2023}, and PynPoint\cite{Amara2012, Stolker2019}, have been widely used for direct imaging of exoplanets.
We benchmark our package, \texttt{torchKLIP}, specifically against \texttt{pyKLIP} to assess its performance.
By leveraging ML/DL approaches in the future, this tool seeks to improve PSF subtraction beyond current methods, with the ultimate goal of achieving contrast ratios as deep as $10^{-10}$ with future instrumentation.

\section{Methodology}\label{methodology}

\subsection{Observing Strategy: Angular Differential Imaging}

\begin{figure}[ht]
    \centering
\includegraphics[width=1\linewidth]{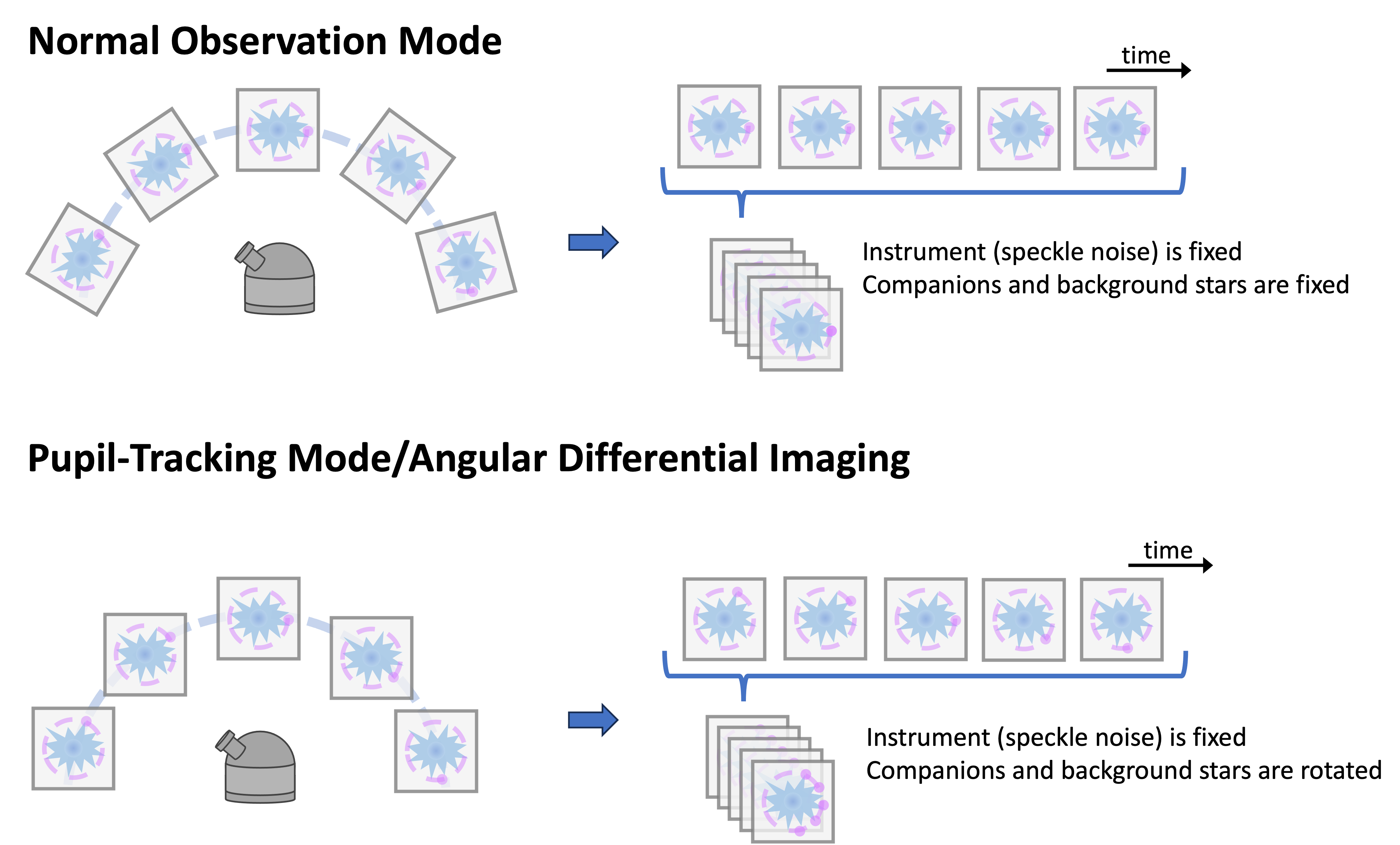}
    \vspace{0.8em}
    \caption{
    A cartoon illustrating the pupil tracking mode and ADI, inspired by Figure 1.3 of Ref.~\citenum{Johnson-Groh2017}. The blue irregular shapes represent instrument speckle noise, while the purple circle represents the faint companion we aim to detect. In the normal observation mode (top), the telescope rotates along with the sky, so all images are aligned throughout the observation. In the pupil tracking mode (bottom), the instrument remains fixed relative to the detector as the sky rotates, causing the companion to rotate around the image center over time.
    }
    \label{fig:ADI}
\end{figure}

One effective observational strategy is angular differential imaging (ADI)\cite{Marois2006}, which utilizes pupil tracking mode (see Figure \ref{fig:ADI}). In this mode, the telescope's pupil (the aperture of the telescope) is kept fixed relative to the detector as the sky rotates. This setup causes the star and any potential companions to rotate around the image center—typically the host star's position on the detector—due to the Earth's rotation, while speckle noise--internal to the instrument--remains stable. This angular diversity leverages the sky's rotation to differentiate between quasi-static speckles and moving planetary signals, thereby enhancing the detection of faint companions.

\subsection{Data Post-processing: PCA-based PSF-subtraction}

\begin{figure}[ht]
    \centering
\includegraphics[width=0.9\linewidth]{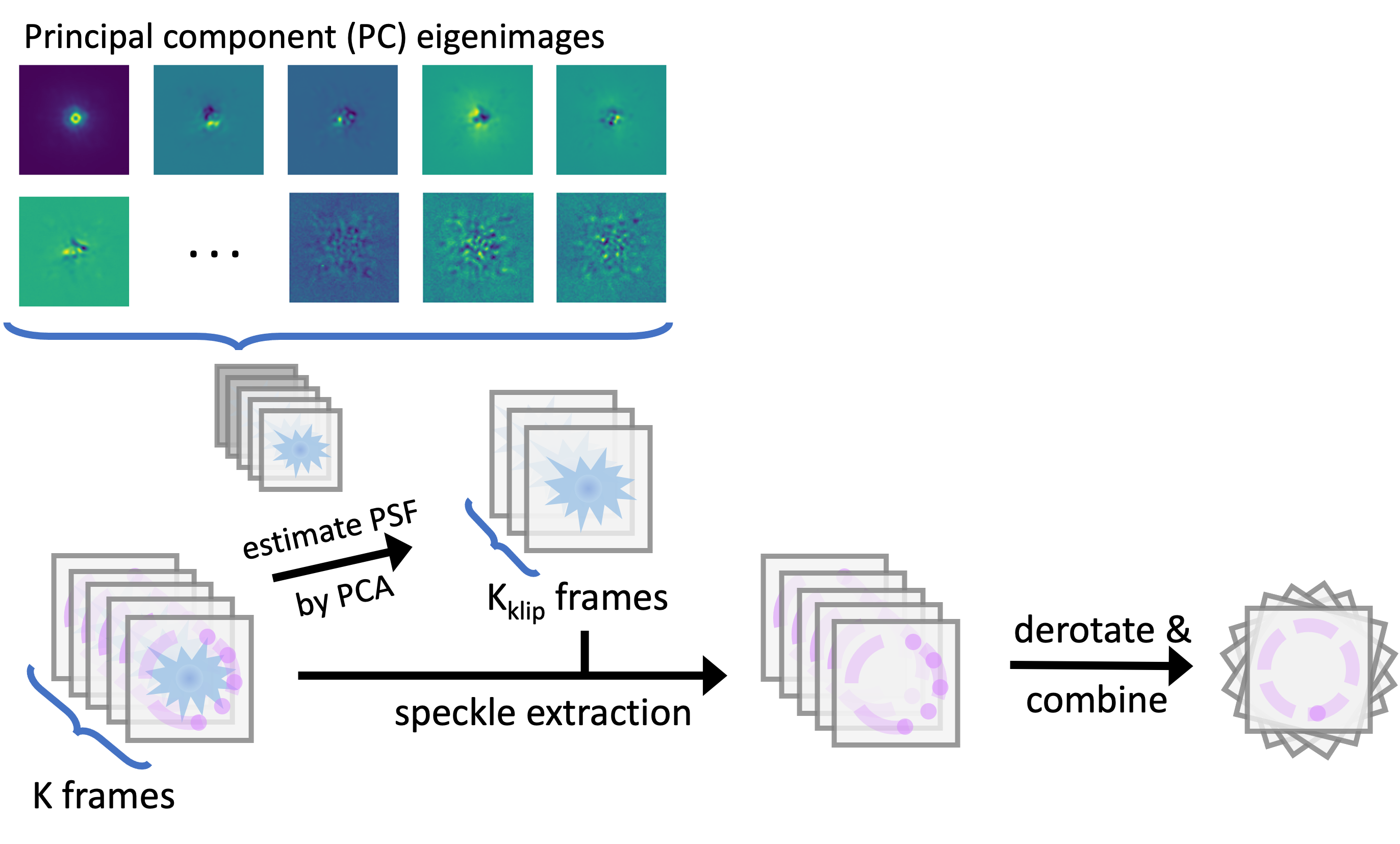}
    \caption{A cartoon illustrating the KLIP algorithm, inspired by Figure 1 of Ref. \citenum{GebBonQuaSch2020}. The irregular blue patterns represent stellar speckle noise from the instrument, while the purple circle marks the location of the faint companion we are trying to detect. The KLIP algorithm works by estimating the stellar PSF using the most significant $K_{\text{klip}}$ eigenimages identified through PCA. After subtracting the PSF from the original dataset, the images are derotated and stacked to align the planetary signal, thereby enhancing the detection of the companion.
    }
    \label{fig:KLIP}
\end{figure}

The commonly used method for data post-processing is KLIP \cite{Soummer2012}, a PCA-based approach (see Figure \ref{fig:KLIP}). PCA identifies $K$ eigenimages from $K$ noise-contaminated observations and truncates the eigenimages to reduce noise. Typically, only the first $K_{\text{klip}} < K$ eigenimages are informative, as they capture the most significant features, while the remaining eigenimages primarily contain noise. The stellar PSF is estimated using these first $K_{\text{klip}}$ eigenimages. After estimating the stellar PSF, it is subtracted from the original dataset to isolate the companions in each frame. The images are then derotated and stacked to enhance the planetary signal. The success of planet detection is evaluated by calculating the signal-to-noise ratio (SNR). 

When applied to a dataset in three dimensions, each frame represents a 2D image captured at a specific time in the sequence.  This collection of frames is denoted as our observed frames $T$. $T$ represents the combined effect of the PSF intensity pattern $I_{\psi}$ (i.e., the response of a telescope instrument to a point source) and the astronomical signal $A$ (i.e., exoplanet). Essentially, $T = I_{\psi} + A$. By accurately estimating $I_{\psi}$, we can extract the astronomical signal $A$ by $A = T - \hat{I_{\psi}}$, where $\hat{I}_{\psi}$ is our estimated PSF intensity pattern. 

The implementation of \texttt{torchKLIP} follows the implementation described in Ref.~\citenum{Soummer2012} and detailed by Refs.~\citenum{Long2020} and \citenum{Long2021}. The code will be made available at \url{https://github.com/astrochialinko/torchKLIP}.

\subsection{Test Cases}

\begin{figure}[ht]
    \centering
\includegraphics[width=0.83\linewidth]{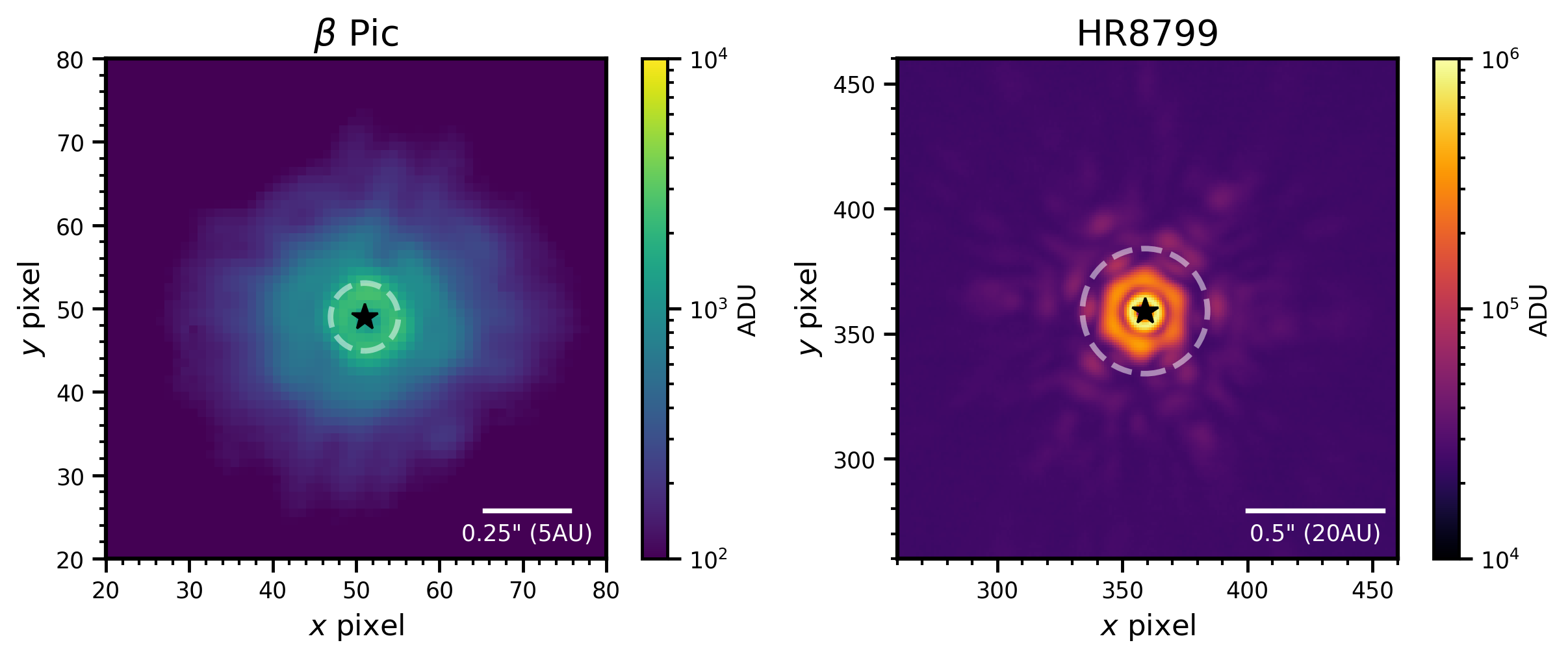}
    \caption{A representative frame of $\beta$ Pic (left) and HR8799 (right) before data post-processing. Stellar speckle noise is clearly visible in both figures and represents the primary noise source that the post-processing aims to mitigate in order to detect faint planets. The white dashed lines indicate the regions masked for post-processing: an inner radius of 4 pixels for $\beta$ Pic and 25 pixels for HR8799. The black star denotes the position of the host star. The color bars use a logarithmic scale and are in units of analog-to-digital units (ADU) or counts.
    }
    \label{fig:orginal_data}
\end{figure}

To benchmark our package torchKLIP against the existing package \texttt{pyKLIP} \cite{Wang2015}, we utilized two well-known datasets: Beta Pictoris\cite{Absil2013} and HR8799\cite{Thompson2023} datasets.
The Beta Pictoris (hereafter $\beta$ Pic; $d\sim19.4$ pc) dataset, observed with VLT/NACO in the $L'$-band, consists of 61 frames in a time sequence, each having 2D dimensions of 100 pixels by 100 pixels and including one confirmed planet. The HR8799 ($d\sim40$ pc) dataset, observed with Keck/NIRC2 in the L-band, comprises 124 frames in a time sequence, each with 2D dimensions of 720 x 720 pixels and containing four confirmed planets.
Details of the datasets are listed in Table \ref{tab:dataset}.
Figure \ref{fig:orginal_data} shows single frames from both datasets, where the stellar speckle noise from the instrument is clearly visible.

\begin{table*}[h]
\caption{Details of the datasets used for benchmarking \texttt{torchKLIP} and \texttt{pyKLIP}}
\label{tab:dataset}

\begin{threeparttable}
\centering
\makebox[1\linewidth]{\normalsize
\begin{tabular}{lll}
\hline\hline  \noalign {\smallskip}
\textbf{Test Cases} & Beta Pictoris & HR 8799   \\ \noalign {\smallskip}
\hline \noalign {\smallskip}
Filter (Central Wavelength) &  $L'$ (3.8 $\mu$m) &  $L'$ (3.8 $\mu$m)   \\ \noalign {\smallskip}
Observation Date &  2013-02-01 & 2021-07-09     \\ \noalign {\smallskip}
On-source Integration Time & 114 mins  &  65 mins    \\ \noalign {\smallskip}
Instrument &  VLT/NACO & Keck/NIRC2    \\ \noalign {\smallskip}
Coronagraph &  AGPM &  None    \\ \noalign {\smallskip}
Field Rotation  &  83.3$^{\circ}$ &  173$^{\circ}$   \\ \noalign {\smallskip}
Size (frames, y pixels,  x pixels) & (61, 100, 100)  &  (124, 720, 720)    \\ \noalign {\smallskip}
Reference &  Ref.~\citenum{Absil2013}; link to data\tnote{a} & Ref.~\citenum{Thompson2023}; link to data\tnote{b} \\  \noalign {\smallskip}
\hline \noalign {\smallskip}
\end{tabular}
}
\begin{tablenotes}
        \footnotesize
        \item[a] \url{https://github.com/carlos-gg/VIP_extras/tree/master/datasets}
        \item[b] \url{https://vmcatcopy.ipac.caltech.edu/ssw/hands-on/Project_Materials_2021}
    \end{tablenotes}
\end{threeparttable}
\end{table*}

\section{Results}

\begin{figure}[ht]
    \centering
\includegraphics[width=0.86\linewidth]{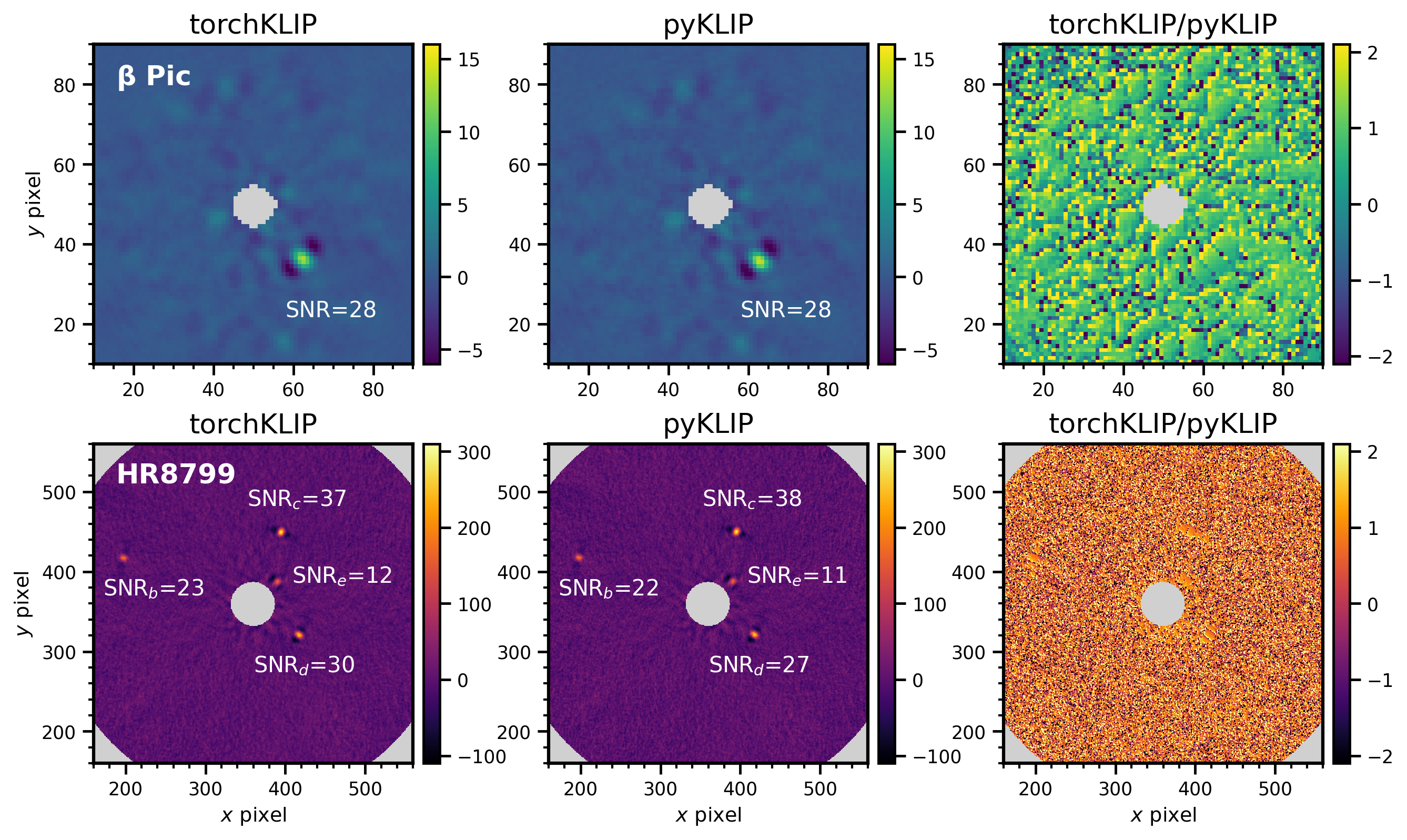}
    \vspace{0.8em}
    \caption{
    Comparison of \texttt{torchKLIP} and \texttt{pyKLIP} package after data processing, including PSF subtraction and combination of derotated images. 
    The upper panel shows images of $\beta$ Pic, while the lower panel displays images of HR8799. In both panels, the left images were processed using \texttt{torchKLIP}, and the middle images were produced using \texttt{pyKLIP}. The same $k_{\text{KLIP}}$ values (9 for $\beta$ Pic and 20 for HR8799) are used to ensure a fair comparison.
    The color bars represent the units of ADU for both the left and middle figures. The right images show the ratio differences between \texttt{torchKLIP} and \texttt{pyKLIP}. The SNR for each detected planet is labeled in the corresponding figures. The gray regions indicate masked areas. The figure demonstrates that the results processed by \texttt{torchKLIP} are comparable with those from \texttt{pyKLIP}.
    }
    \label{fig:compared}
\end{figure}

\begin{figure}[ht]
    \centering
\includegraphics[width=0.83\linewidth]{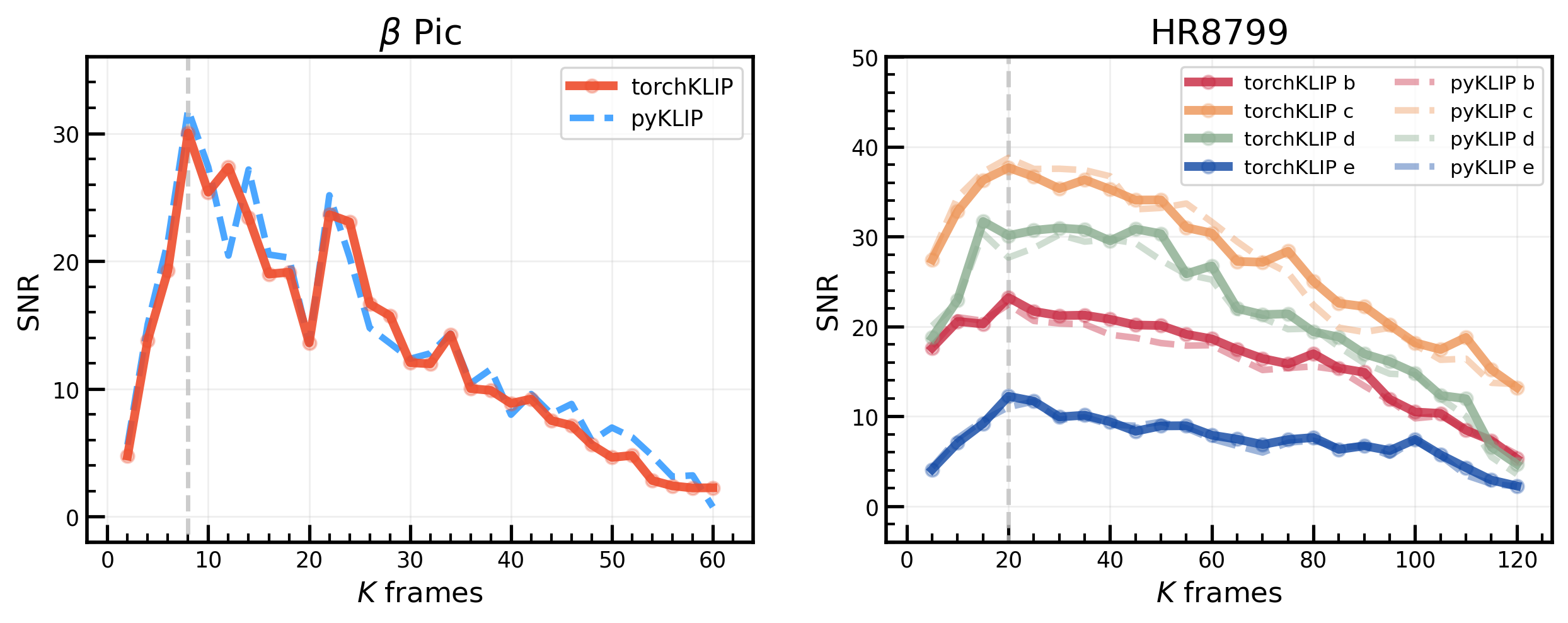}
    \caption{Comparison between SNR and the number of frames ($K$) for \texttt{torchKLIP} (solid lines) and \texttt{pyKLIP} (dashed lines). The left panel shows results for $\beta$ Pic, while the right panel shows results for HR8799, with different color lines representing the four different planets. The vertical gray dashed lines indicate the frames where the maximal SNR is reached (9 for $\beta$ Pic; 20 for HR8799). The figure suggests that \texttt{torchKLIP} and \texttt{pyKLIP} achieve comparable peak SNRs.
}
    \label{fig:snr}
\end{figure}

To benchmark our package \texttt{torchKLIP} against \texttt{pyKLIP}, we compared the SNR and computation time using full-frame PCA on the $\beta$ Pic and HR8799 datasets.
Figure \ref{fig:compared} presents the outputs of \texttt{torchKLIP} and \texttt{pyKLIP} on both datasets using the same $k_{\text{KLIP}}$ values to ensure a fair comparison. The images generated by both packages display similar SNRs for the detected planets, and the ratio differences between \texttt{torchKLIP} and \texttt{pyKLIP} further confirm this consistency, suggesting that both algorithms effectively mitigate residual speckle noise.
Figure \ref{fig:snr} shows how the SNR varies with different $k_{\text{KLIP}}$ values for both \texttt{torchKLIP} and \texttt{pyKLIP} in the $\beta$ Pic and HR8799 datasets. Both algorithms achieve the maximum SNR at $k_{\text{KLIP}} = 9$ for $\beta$ Pic and $k_{\text{KLIP}} = 20$ for HR8799, with consistent SNRs across the variation in $k_{\text{KLIP}}$. These results further suggest that \texttt{torchKLIP} and \texttt{pyKLIP} are equally effective in mitigating residual speckle noise.

\begin{figure}[ht]
    \centering
\includegraphics[width=0.83\linewidth]{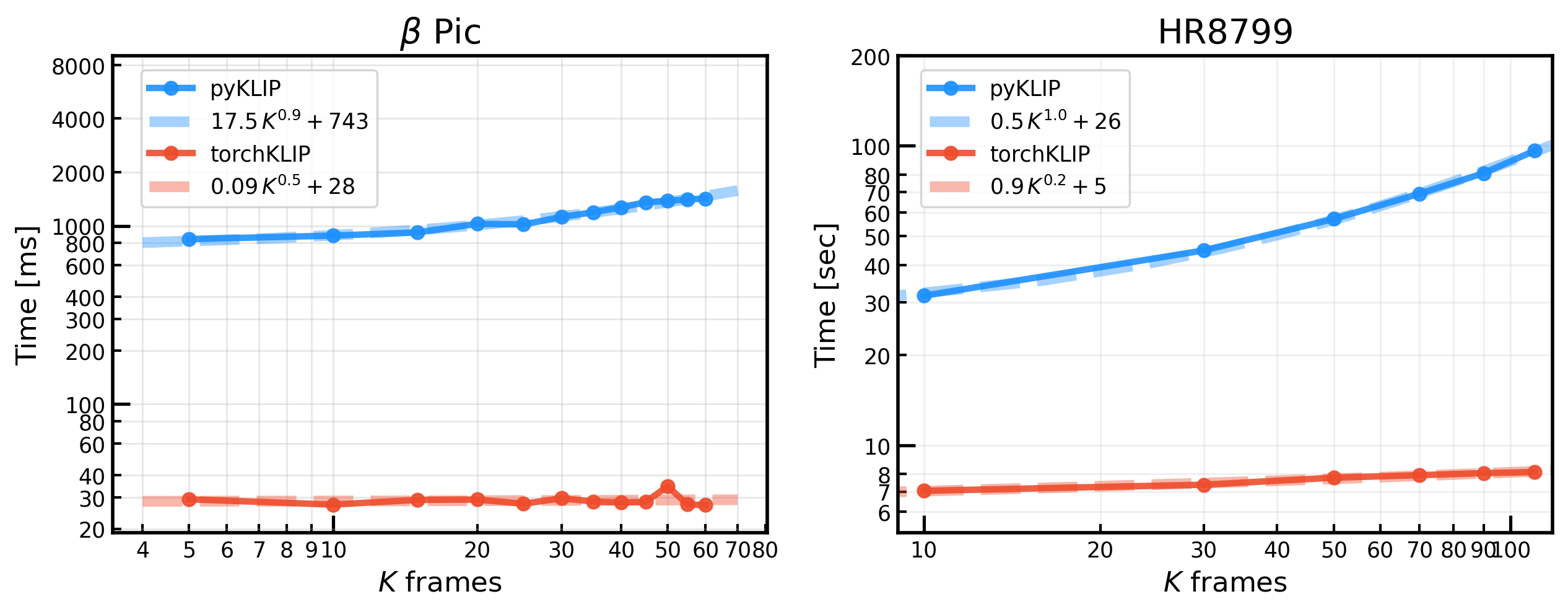}
    \caption{Comparison of computation time between \texttt{torchKLIP} and \texttt{pyKLIP} as the number of $K$ frames increases. The left plot corresponds to $\beta$ Pic, and the right plot to HR8799. Computation times for \texttt{torchKLIP} and \texttt{pyKLIP} are indicated by the solid orange and blue lines, respectively. The dashed lines represent the fitted curves for each method. Overall, \texttt{torchKLIP} consistently shows shorter computation times than \texttt{pyKLIP}, with a lower rate of increase as the number of $K$ frames grows.
    }
    \label{fig:time}
\end{figure}

In Figure \ref{fig:time}, we compared the computation times of \texttt{torchKLIP} and \texttt{pyKLIP} as the number of $k_{\text{KLIP}}$ frames increases for both the $\beta$ Pic and HR8799 datasets. These measurements were taken on an Intel Core i7-11700 CPU (16 cores, 32 GB RAM) using 16 cores. The results demonstrate that \texttt{torchKLIP} consistently outperforms \texttt{pyKLIP} in terms of computation speed, with significantly shorter processing times across all frame numbers. Additionally, the rate at which computation time increases with the number of frames is notably slower for \texttt{torchKLIP}, indicating superior scalability and efficiency, especially as the dataset size grows. 
It is worth noting that \texttt{torchKLIP} is currently a prototype, performing only full-frame PCA, while \texttt{pyKLIP} includes additional features beyond full-frame PCA, such as annuli and subsections. These extra features in \texttt{pyKLIP} can potentially achieve higher SNR results, but they may also contribute to longer processing times. To ensure a fair comparison of both SNR and computation time, we implemented full-frame PCA in both \texttt{torchKLIP} and \texttt{pyKLIP} for this benchmark.
These comparisons show that \texttt{torchKLIP} performs comparably with \texttt{pyKLIP} in terms of SNR while offering advantages in computational efficiency.

\section{Summary}

We developed, \texttt{torchKLIP}, a KLIP implementation for PSF subtraction within the \texttt{PyTorch} framework, which showed negligible differences or improvements in SNR compared to the existing package, \texttt{pyKLIP}. As for the computing time, \texttt{torchKLIP} demonstrated shorter computation times and a lower increase rate in computation time as the number of $k$ frames increased. Future work will focus on expanding comparisons to include additional parameters beyond full-frame PCA, such as annuli and subsections. Additionally, efforts will be made to extend \texttt{torchKLIP}'s applicability to more instruments and observational strategies, including Reference Differential Imaging (RDI)\cite{Lafreniere2007} and Spectral Differential Imaging (SDI)\cite{Vigan2010}. The integration of machine learning and deep learning algorithms into \texttt{torchKLIP} using simulations and data from high-contrast imaging testbeds is also planned.

\acknowledgments 
 
Portions of this research were supported by funding from generous anonymous philanthropic donations to the Steward Observatory of the College of Science at the University of Arizona.
CLK thanks the Code/Astro Workshop for providing valuable training in the development of open-source software packages

\bibliographystyle{spiebib} 

\begin{thebibliography}{10}

\bibitem{Fischer2014}
{Fischer}, D.~A., {Howard}, A.~W., {Laughlin}, G.~P., {Macintosh}, B., {Mahadevan}, S., {Sahlmann}, J., and {Yee}, J.~C., ``{Exoplanet Detection Techniques},'' in [{\em Protostars and Planets VI}{\nolinebreak\hspace{0.1em}]},  {Beuther}, H., {Klessen}, R.~S., {Dullemond}, C.~P., and {Henning}, T., eds.,  715--737 (Jan. 2014).

\bibitem{Currie2023}
{Currie}, T., {Biller}, B., {Lagrange}, A., {Marois}, C., {Guyon}, O., {Nielsen}, E.~L., {Bonnefoy}, M., and {De Rosa}, R.~J., ``{Direct Imaging and Spectroscopy of Extrasolar Planets},'' in [{\em Protostars and Planets VII}{\nolinebreak\hspace{0.1em}]},  {Inutsuka}, S., {Aikawa}, Y., {Muto}, T., {Tomida}, K., and {Tamura}, M., eds., {\em Astronomical Society of the Pacific Conference Series} {\bf 534},  799 (July 2023).

\bibitem{Follette2023}
{Follette}, K.~B., ``{An Introduction to High Contrast Differential Imaging of Exoplanets and Disks},'' {\em \pasp}~{\bf 135},  093001 (Sept. 2023).

\bibitem{Milli2016}
{Milli}, J., {Mawet}, D., {Mouillet}, D., {Kasper}, M., and {Girard}, J.~H., ``{Adaptive Optics in High-Contrast Imaging},'' in [{\em Astronomy at High Angular Resolution}{\nolinebreak\hspace{0.1em}]},  {Boffin}, H. M.~J., {Hussain}, G., {Berger}, J.-P., and {Schmidtobreick}, L., eds., {\em Astrophysics and Space Science Library} {\bf 439},  17 (Jan. 2016).

\bibitem{Soummer2012}
{Soummer}, R., {Pueyo}, L., and {Larkin}, J., ``{Detection and Characterization of Exoplanets and Disks Using Projections on Karhunen-Lo{\`e}ve Eigenimages},'' {\em \apjl}~{\bf 755},  L28 (Aug. 2012).

\bibitem{Jolliffe2016}
Jolliffe, I.~T. and Cadima, J., ``Principal component analysis: a review and recent developments,'' {\em Philosophical Transactions of the Royal Society A: Mathematical, Physical and Engineering Sciences}~{\bf 374} (2016).

\bibitem{Guyon2005}
{Guyon}, O., ``{Limits of Adaptive Optics for High-Contrast Imaging},'' {\em \apj}~{\bf 629},  592--614 (Aug. 2005).

\bibitem{Gaudi2020}
{Gaudi}, B.~S., {Seager}, S., {Mennesson}, B., {Kiessling}, A., {Warfield}, K., {Cahoy}, K., {Clarke}, J.~T., {Domagal-Goldman}, S., {Feinberg}, L., {Guyon}, O., {Kasdin}, J., {Mawet}, D., {Plavchan}, P., {Robinson}, T., {Rogers}, L., {Scowen}, P., {Somerville}, R., {Stapelfeldt}, K., {Stark}, C., {Stern}, D., {Turnbull}, M., {Amini}, R., {Kuan}, G., {Martin}, S., {Morgan}, R., {Redding}, D., {Stahl}, H.~P., {Webb}, R., {Alvarez-Salazar}, O., {Arnold}, W.~L., {Arya}, M., {Balasubramanian}, B., {Baysinger}, M., {Bell}, R., {Below}, C., {Benson}, J., {Blais}, L., {Booth}, J., {Bourgeois}, R., {Bradford}, C., {Brewer}, A., {Brooks}, T., {Cady}, E., {Caldwell}, M., {Calvet}, R., {Carr}, S., {Chan}, D., {Cormarkovic}, V., {Coste}, K., {Cox}, C., {Danner}, R., {Davis}, J., {Dewell}, L., {Dorsett}, L., {Dunn}, D., {East}, M., {Effinger}, M., {Eng}, R., {Freebury}, G., {Garcia}, J., {Gaskin}, J., {Greene}, S., {Hennessy}, J., {Hilgemann}, E., {Hood}, B., {Holota}, W., {Howe}, S., {Huang}, P., {Hull}, T., {Hunt}, R.,
  {Hurd}, K., {Johnson}, S., {Kissil}, A., {Knight}, B., {Kolenz}, D., {Kraus}, O., {Krist}, J., {Li}, M., {Lisman}, D., {Mandic}, M., {Mann}, J., {Marchen}, L., {Marrese-Reading}, C., {McCready}, J., {McGown}, J., {Missun}, J., {Miyaguchi}, A., {Moore}, B., {Nemati}, B., {Nikzad}, S., {Nissen}, J., {Novicki}, M., {Perrine}, T., {Pineda}, C., {Polanco}, O., {Putnam}, D., {Qureshi}, A., {Richards}, M., {Eldorado Riggs}, A.~J., {Rodgers}, M., {Rud}, M., {Saini}, N., {Scalisi}, D., {Scharf}, D., {Schulz}, K., {Serabyn}, G., {Sigrist}, N., {Sikkia}, G., {Singleton}, A., {Shaklan}, S., {Smith}, S., {Southerd}, B., {Stahl}, M., {Steeves}, J., {Sturges}, B., {Sullivan}, C., {Tang}, H., {Taras}, N., {Tesch}, J., {Therrell}, M., {Tseng}, H., {Valente}, M., {Van Buren}, D., {Villalvazo}, J., {Warwick}, S., {Webb}, D., {Westerhoff}, T., {Wofford}, R., {Wu}, G., {Woo}, J., {Wood}, M., {Ziemer}, J., {Arney}, G., {Anderson}, J., {Ma{\'\i}z-Apell{\'a}niz}, J., {Bartlett}, J., {Belikov}, R., {Bendek}, E., {Cenko}, B.,
  {Douglas}, E., {Dulz}, S., {Evans}, C., {Faramaz}, V., {Feng}, Y.~K., {Ferguson}, H., {Follette}, K., {Ford}, S., {Garc{\'\i}a}, M., {Geha}, M., {Gelino}, D., {G{\"o}tberg}, Y., {Hildebrandt}, S., {Hu}, R., {Jahnke}, K., {Kennedy}, G., {Kreidberg}, L., {Isella}, A., {Lopez}, E., {Marchis}, F., {Macri}, L., {Marley}, M., {Matzko}, W., {Mazoyer}, J., {McCandliss}, S., {Meshkat}, T., {Mordasini}, C., {Morris}, P., {Nielsen}, E., {Newman}, P., {Petigura}, E., {Postman}, M., {Reines}, A., {Roberge}, A., {Roederer}, I., {Ruane}, G., {Schwieterman}, E., {Sirbu}, D., {Spalding}, C., {Teplitz}, H., {Tumlinson}, J., {Turner}, N., {Werk}, J., {Wofford}, A., {Wyatt}, M., {Young}, A., and {Zellem}, R., ``{The Habitable Exoplanet Observatory (HabEx) Mission Concept Study Final Report},'' {\em arXiv e-prints} ,  arXiv:2001.06683 (Jan. 2020).

\bibitem{Cantalloube2020}
Cantalloube, F., Gomez-Gonzalez, C., Absil, O., Cantero, C., Bacher, R., Bonse, M.~J., Bottom, M., Dahlqvist, C.-H., Desgrange, C., Flasseur, O., Fuhrmann, T., Henning, T., Jensen-Clem, R., Kenworthy, M., Mawet, D., Mesa, D., Meshkat, T., Mouillet, D., M{\"u}ller, A., Nasedkin, E., Pairet, B., Pi{\'e}rard, S., Ruffio, J.-B., Samland, M., Stone, J., and Droogenbroeck, M.~V., ``{Exoplanet imaging data challenge: benchmarking the various image processing methods for exoplanet detection},'' in [{\em Adaptive Optics Systems VII}{\nolinebreak\hspace{0.1em}]},  Schreiber, L., Schmidt, D., and Vernet, E., eds.,  {\bf 11448},  114485A, International Society for Optics and Photonics, SPIE (2020).

\bibitem{Gebhard2022}
{Gebhard}, T.~D., {Bonse}, M.~J., {Quanz}, S.~P., and {Sch{\"o}lkopf}, B., ``{Half-sibling regression meets exoplanet imaging: PSF modeling and subtraction using a flexible, domain knowledge-driven, causal framework},'' {\em \aap}~{\bf 666},  A9 (Oct. 2022).

\bibitem{Flasseur2024}
{Flasseur}, O., {Bodrito}, T., {Mairal}, J., {Ponce}, J., {Langlois}, M., and {Lagrange}, A.-M., ``{deep PACO: combining statistical models with deep learning for exoplanet detection and characterization in direct imaging at high contrast},'' {\em \mnras}~{\bf 527},  1534--1562 (Jan. 2024).

\bibitem{Bonse2024}
{Bonse}, M.~J., {Gebhard}, T.~D., {Dannert}, F.~A., {Absil}, O., {Cantalloube}, F., {Christiaens}, V., {Cugno}, G., {Garvin}, E.~O., {Hayoz}, J., {Kasper}, M., {Matthews}, E., {Sch{\"o}lkopf}, B., and {Quanz}, S.~P., ``{Use the 4S (Signal-Safe Speckle Subtraction): Explainable Machine Learning reveals the Giant Exoplanet AF Lep b in High-Contrast Imaging Data from 2011},'' {\em arXiv e-prints} ,  arXiv:2406.01809 (June 2024).

\bibitem{Paszke2019}
Paszke, A., Gross, S., Massa, F., Lerer, A., Bradbury, J., Chanan, G., Killeen, T., Lin, Z., Gimelshein, N., Antiga, L., Desmaison, A., Kopf, A., Yang, E., DeVito, Z., Raison, M., Tejani, A., Chilamkurthy, S., Steiner, B., Fang, L., Bai, J., and Chintala, S., ``Pytorch: An imperative style, high-performance deep learning library,'' in [{\em Advances in Neural Information Processing Systems 32}{\nolinebreak\hspace{0.1em}]},   8024--8035, Curran Associates, Inc. (2019).

\bibitem{Wang2015}
{Wang}, J.~J., {Ruffio}, J.-B., {De Rosa}, R.~J., {Aguilar}, J., {Wolff}, S.~G., and {Pueyo}, L., ``{pyKLIP: PSF Subtraction for Exoplanets and Disks}.'' Astrophysics Source Code Library, record ascl:1506.001 (June 2015).

\bibitem{Gonzalez2017}
{Gomez Gonzalez}, C.~A., {Wertz}, O., {Absil}, O., {Christiaens}, V., {Defr{\`e}re}, D., {Mawet}, D., {Milli}, J., {Absil}, P.-A., {Van Droogenbroeck}, M., {Cantalloube}, F., {Hinz}, P.~M., {Skemer}, A.~J., {Karlsson}, M., and {Surdej}, J., ``{VIP: Vortex Image Processing Package for High-contrast Direct Imaging},'' {\em \aj}~{\bf 154},  7 (July 2017).

\bibitem{Christiaens2023}
{Christiaens}, V., {Gonzalez}, C., {Farkas}, R., {Dahlqvist}, C.-H., {Nasedkin}, E., {Milli}, J., {Absil}, O., {Ngo}, H., {Cantero}, C., {Rainot}, A., {Hammond}, I., {Bonse}, M., {Cantalloube}, F., {Vigan}, A., {Kompella}, V., and {Hancock}, P., ``{VIP: A Python package for high-contrast imaging},'' {\em The Journal of Open Source Software}~{\bf 8},  4774 (Jan. 2023).

\bibitem{Amara2012}
{Amara}, A. and {Quanz}, S.~P., ``{PYNPOINT: an image processing package for finding exoplanets},'' {\em \mnras}~{\bf 427},  948--955 (Dec. 2012).

\bibitem{Stolker2019}
{Stolker}, T., {Bonse}, M.~J., {Quanz}, S.~P., {Amara}, A., {Cugno}, G., {Bohn}, A.~J., and {Boehle}, A., ``{PynPoint: a modular pipeline architecture for processing and analysis of high-contrast imaging data},'' {\em \aap}~{\bf 621},  A59 (Jan. 2019).

\bibitem{Marois2006}
{Marois}, C., {Lafreni{\`e}re}, D., {Doyon}, R., {Macintosh}, B., and {Nadeau}, D., ``{Angular Differential Imaging: A Powerful High-Contrast Imaging Technique},'' {\em \apj}~{\bf 641},  556--564 (Apr. 2006).

\bibitem{Johnson-Groh2017}
{Johnson-Groh}, M., ``{From Exoplanets to Quasars: Adventures in Angular Differential Imaging},'' {\em arXiv e-prints} ,  arXiv:1704.00317 (Apr. 2017).

\bibitem{Long2020}
Long, J., ``Finding a planet by subtracting starlight.'' September 28 2020, \url{https://joseph-long.com/writing/klip-in-numpy}.
\newblock (Accessed: 25 August 2024).

\bibitem{Long2021}
{Long}, J.~D. and {Males}, J.~R., ``{Unlocking Starlight Subtraction in Full-data-rate Exoplanet Imaging by Efficiently Updating Karhunen-Lo{\`e}ve Eigenimages},'' {\em \aj}~{\bf 161},  166 (Apr. 2021).

\bibitem{GebBonQuaSch2020}
Gebhard, T.~D., Bonse, M.~J., Quanz, S.~P., and Sch{\"o}lkopf, B., ``Physically constrained causal noise models for high-contrast imaging of exoplanets,'' in [{\em Machine Learning and the Physical Sciences - Workshop at the 34th Conference on Neural Information Processing Systems (NeurIPS)}{\nolinebreak\hspace{0.1em}]},  (2020).

\bibitem{Absil2013}
{Absil}, O., {Milli}, J., {Mawet}, D., {Lagrange}, A.~M., {Girard}, J., {Chauvin}, G., {Boccaletti}, A., {Delacroix}, C., and {Surdej}, J., ``{Searching for companions down to 2 AU from {\ensuremath{\beta}} Pictoris using the L'-band AGPM coronagraph on VLT/NACO},'' {\em \aap}~{\bf 559},  L12 (Nov. 2013).

\bibitem{Thompson2023}
{Thompson}, W., {Marois}, C., {Do {\'O}}, C.~R., {Konopacky}, Q., {Ruffio}, J.-B., {Wang}, J., {Skemer}, A.~J., {De Rosa}, R.~J., and {Macintosh}, B., ``{Deep Orbital Search for Additional Planets in the HR 8799 System},'' {\em \aj}~{\bf 165},  29 (Jan. 2023).

\bibitem{Lafreniere2007}
{Lafreni{\`e}re}, D., {Marois}, C., {Doyon}, R., {Nadeau}, D., and {Artigau}, {\'E}., ``{A New Algorithm for Point-Spread Function Subtraction in High-Contrast Imaging: A Demonstration with Angular Differential Imaging},'' {\em \apj}~{\bf 660},  770--780 (May 2007).

\bibitem{Vigan2010}
{Vigan}, A., {Moutou}, C., {Langlois}, M., {Allard}, F., {Boccaletti}, A., {Carbillet}, M., {Mouillet}, D., and {Smith}, I., ``{Photometric characterization of exoplanets using angular and spectral differential imaging},'' {\em \mnras}~{\bf 407},  71--82 (Sept. 2010).

\end{thebibliography}

\end{document}